\documentclass{article}
\pdfoutput=1
\usepackage{nips15submit_e,times}
\usepackage{hyperref}
\usepackage{url}
\usepackage{graphicx}
\usepackage{caption}
\usepackage{subcaption}
\usepackage{float}
\usepackage{amsmath}
\begin{document}
\title{A Novel Ensemble Deep Learning Model for Stock Prediction Based on Stock Prices and News}

\author{
Yang Li\\
Department of Computer Science\\
Georgia State University\\
Atlanta, GA 30303 \\
\texttt{yli48@gsu.edu} 
\And
Yi Pan\\
Department of Computer Science\\
Georgia State University\\
Atlanta, GA 30303 \\
\texttt{yipan@gsu.edu} 
}

\nipsfinalcopy

\maketitle

\begin{abstract}
In recent years, machine learning and deep learning have become popular methods for financial data analysis, including financial textual data, numerical data, and graphical data. Stock Investors make investment decisions according to various factors such as company’s charts, market indices, and textual information, e.g., newspapers, news-carrying micro-blogs, or other miscellaneous sources\cite{10.1093/rfs/hhl024}. This paper proposes a novel deep learning approach to predict future stock movement. The model employs a blending ensemble learning method to combine two recurrent neural networks (RNNs) followed by a fully connected neural network (FCNN). In our research, we use the S\&P 500 Index as our test case. Our experiments show that our blending ensemble deep learning model outperforms the best existing prediction model substantially using the same dataset, reducing the mean squared error (MSE) from 438.94 to 186.32, a 57.55\% reduction, increasing precision rate by 40\%, Recall by 50\%,  F1-Score by 44.78\%, and Movement Direction Accuracy (MDA) by 33.34\%, respectively. The purpose of this work is to explain our design philosophy and show that ensemble deep learning technologies can truly predict future stock price trends more effectively and can better assist investors in making the right investment decision than other traditional methods. 
\end{abstract}

\section{Introduction}
Many factors may affect stock prices in various ways. The stock prices change by market forces, which means the stock price changes react to supply and demand in the stock market. If more people want to buy a stock (demand) than sell it (supply), then the price moves up. Similarly, if more people want to sell a stock than buy it, there would be greater supply than demand, and the price would fall. Stock supply and demand are affected by many things. Supply factors include company share issues (e.g., releases new shares to the public), share buybacks (e.g., a company buys back its own shares from investors to reduce supply) and sellers (e.g., the investors responsible for pushing shares back into the market, increasing the supply). Demand factors include company news (e.g., a new product launch, missed targets, good performance), economic factors (e.g., interest rate changes), industry trends (e.g., a booming industry), market sentiment (could be psychological and subjective) and unexpected events (e.g., natural disasters or the death of a government leader). Normally, we can get these supply and demand factors from the financial news, companies' newsletters or their annual reports. For instance, when Apple announces a new product, many people would like to purchase it, and its performance usually would be better soon. Thus more people are interested in the Apple stock, then the Apple stock demand increases, which will lead to a rise in the Apple stock price. On the other hand, when COVID-19 spreads around the world, many airlines cut their flights, and it is expected their performance would be bad in a short term. Thus, more people want to sell airline stocks; then the airline stock supply will rise, and their price will go down. If the price goes up, the quantity demanded goes down (but demand itself stays the same). If the price decreases, quantity demanded increases. This is the Law of Demand. If the quantity demanded decreases, the stock price probably would fall. Also people's sentiment or belief plays a role in determining a stock price. Political situations or international affairs may also affect stock prices. Hence, this is a complicated process among the stock supply, demand, and prices. However, there are a few primary factors that affect the stock supply and demand like company news, company performance, industry performance, investor sentiment (e.g., whether in bull or bear market), and other major economic factors described in \cite{Islam2015FactorsAT}. If we focus on the major factors, and trace back the historical stock prices, we may be able to predict future stock prices quite accurately. People usually have a short memory about stock factors. Hence, determining a suitable historical window size is important to correctly predict stock prices. If the window size is too large compared with human memory, many factors or news are forgotten by investors and obsolete already and the prediction will not be good. On the other hand, if the window is too short compared with human memory, many news or sentiments outside the window are still remain in people's brain, the prediction will also be bad. Hence, a wrong historical window size is detrimental to our successful stock price predictions.

Stock price prediction is a series of continuous predictions since the stock price is constantly changing to react to timely news and announcements. Therefore, it is very challenging for computer scientists to use Artificial Intelligence to predict future stock movements because it is hard for a computer to receive the latest information and respond immediately. Computer Scientists are currently not particularly successful in stock price prediction for several reasons. Most of the previous works\cite{choi2018stock}\cite{Kordonis2016StockPF} often used either textual data like news, twitter, or blogs or numerical data like stock price information instead of using both textual information and statistical information\cite{7550882}. Since the stock price is related to many factors, only considering one or two factors is unable to provide enough information to forecast the stock price trend. Including as much relevant and useful information as possible will guarantee a better prediction.

Furthermore, previous works\cite{choi2018stock}\cite{Kordonis2016StockPF} only use the target company's information on the training model without considering that the target company's competitors or the information of companies in related industries. These types of information will also affect the target company's stock movement. Therefore, the result is not very satisfactory and persuasive because the information provided is insufficient. Moreover, some of the previous works, which used the textual information, did not consider time series. However, the timeline is a significant factor for stock price prediction. 

This paper proposes to use sentiment analysis to extract useful information from multiple textual data sources and a blending ensemble deep learning model to predict future stock movement. The blending ensemble model contains two levels. The first level contains two Recurrent Neural Networks (RNNs), one Long-Short Term Memory network (LSTM) and one Gated Recurrent Units network (GRU), followed by a fully connected neural network as the second level model. The RNNs, LSTM, and GRU models can effectively capture the time series events in the input data, and the fully connected neural network is used to ensemble several individual prediction results to further improve the prediction accuracy.

\section{Related Work}
Three previous works have a significant impact on this research. Last year, Li\cite{li2019dplstm} proposed a novel approach to use differential privacy to robust the LSTM model for stock prediction. The experimental results have shown using differential privacy can robust the LSTM model and improve prediction results. The Differential Privacy-inspired LSTM (DP-LSTM) approach inspired us to attempt to use a different approach to predict stock movements. In their paper “Deep Learning for Stock Prediction Using Numerical and Textual Information" \cite{7550882}, the authors stated that converting newspaper articles into distributed representations via paragraph Vector and modeling the temporal effects of the past events on opening prices about multiple companies with LSTM could increase stock price prediction accuracy. This previous work also suggests using numerical data and textual data as primary sources to predict future stock prices with LSTM. Leonardo Pinheiro and Mark Dras showed that they explored RNNs with character-level language model using pre-training for both intraday and interday stock market forecasting and their technique is competitive with other state-of-the-art approaches\cite{dos-santos-pinheiro-dras-2017-stock}. Our architecture leverages their successful experiences and creates a new model to perform a better prediction.

\section{Preliminaries}
\label{headings}
\subsection{Data Source}
% Data Overview
The data used in this research was obtained from the paper “DP-LSTM: Differential Privacy-inspired LSTM for Stock Prediction Using Financial News"\cite{li2019dplstm}. We taxonomize the data into two categories: news and stock data; the news data are obtained from CNBC.com, Reuters.com, WSJ.com, Fortune.com, and dates range from December 2017 up to the end of June 2018. CNBC is the world leader in business news and has a real-time financial market coverage. Reuters is an international news organization founded in October 1851; it is one of the industry leaders for online information for tax, accounting, and finance news. The Wall Street Journal (WSJ) is one of the largest American business-focused news organization based in New York City. Fortune is an American multinational business magazine. We consider these four financial news data because these are the four most prominent business news organizations and hence the quality of the news there is exceptional.  
% News Data
For the news data, only news articles from the financial domain were considered. Moreover, Ding et al.\cite{ding-etal-2014-using} advised that news titles can provide adequate information to represent news articles and are more helpful for prediction compared to the article's contents. Besides, the article's content might add extra noises to the model that might cause the model to have poor performance, and it is also hard to accurately summarize the article's content using Natural Language Processing (NLP). Hence, we only use the title of the news to extract sentiment scores. 
% Stock Data
The stock data is the S\&P 500 Index with the same date range as the news data. The S\&P 500 Index is a stock market index that measures the stock performance of the 500 largest publicly traded companies in the United States. The S\&P 500 is one of the best representations of the U.S. stock market. Since our experiment uses news data and stock data to predict future stock market movement and prices, we will only use adjusted closing stock price as the target value. Adjusted closing price amends a stock's closing price to accurately reflect that stock's value after accounting for any corporate actions. It is considered to be the true price of that stock and is often used when examining historical returns or performing a detailed analysis of historical returns. 

% Data Processing
\subsection{Data Pre-processing}

The news data are pre-processed with Aware Dictionary and Sentiment Reasoner(VADER) to generate sentiment scores. VADER is a lexicon and rule-based model for general sentiment analysis\cite{Hutto2014VADERAP}. According to Kirlic's research, there is almost no difference between VADER and human decision making\cite{ijrtem_2018}. In addition, VADER not only provides positive and negative scores, but also gives us how positive or negative the sentiment is. Many python libraries include a pre-trained VADER model that makes it very convenient and efficient for us to use. After pre-processing the news data, the VADER will give a compound score; the compound score is a metric that calculates the sum of all the lexicon ratings, which has a normalized value between -1, which represents most extreme negative and +1, which represents most extreme positive\cite{li2019dplstm}. Of course, 0 means neutral news.%example
For example, if the news title is “The Price of the U.S. Dollar is Rising" and its compound score is 0.64, which means this news title is positive, and the positivity weight is about 0.64. The opposite example would be “The Price of the U.S. Dollar is Falling" and if its compound score is -0.56, which means this news title is negative, and the negativity weight is about -0.56.

During the pre-processing, all the null data are removed from the dataset, and all news data and stock data are combined together. Since the stock market only opens and closes during the weekdays, therefore weekends are not included in the dataset. Our dataset contains 121 trading days and a total of six columns; the first column contains the date corresponding to the news data and stock data in the next 5 columns. The news data contains the WSJ news compound score, the Reuter's news compound score, the CBNC news compound score, the Fortune news compound score, and the stock data are the adjusted closing prices.

\begin{figure}[hbt!]
\begin{center}
\includegraphics[scale=0.28]{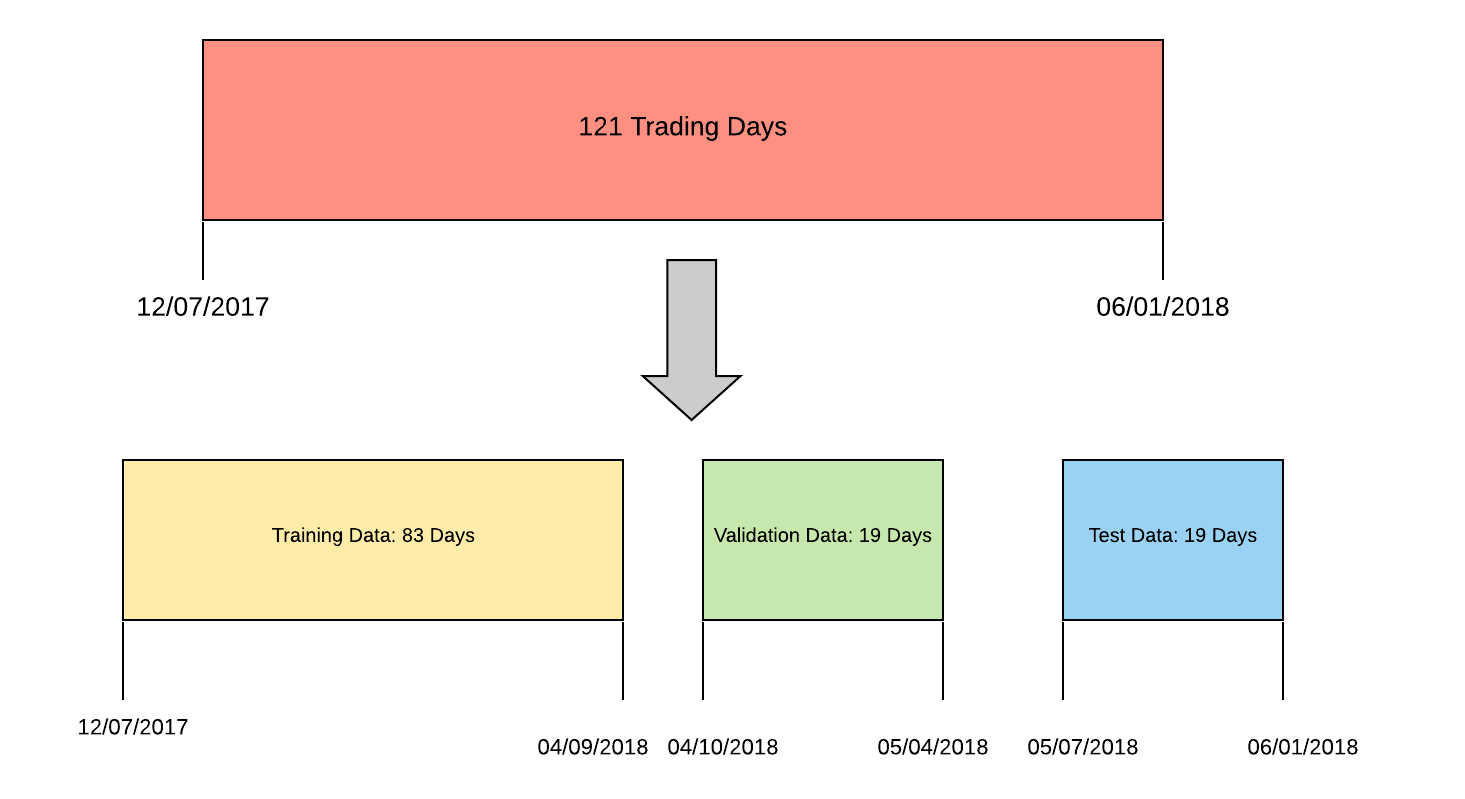}
\caption{\small Schematic diagram of the splited dataset.}
\label{fig:dataset_structure}
\end{center}
\end{figure}

As shown in Figure \ref{fig:dataset_structure}, we have split the data into three parts: training data, validation data, and test data. The training data are those obtained from 12/07/2017 to 04/09/2018; the validation data are those data between 04/10/2018 and 05/04/2018; the test data are from 05/07/2018 to 06/01/2018. The training dataset is used to train the level 0 sub-models, the validation dataset is used to prepare the level 1 model, and the test dataset used to evaluate our prediction performance. The details of the model architecture will be discussed later in section 4.4.

\begin{figure}[hbt!]
\begin{center}
\includegraphics[scale=0.12]{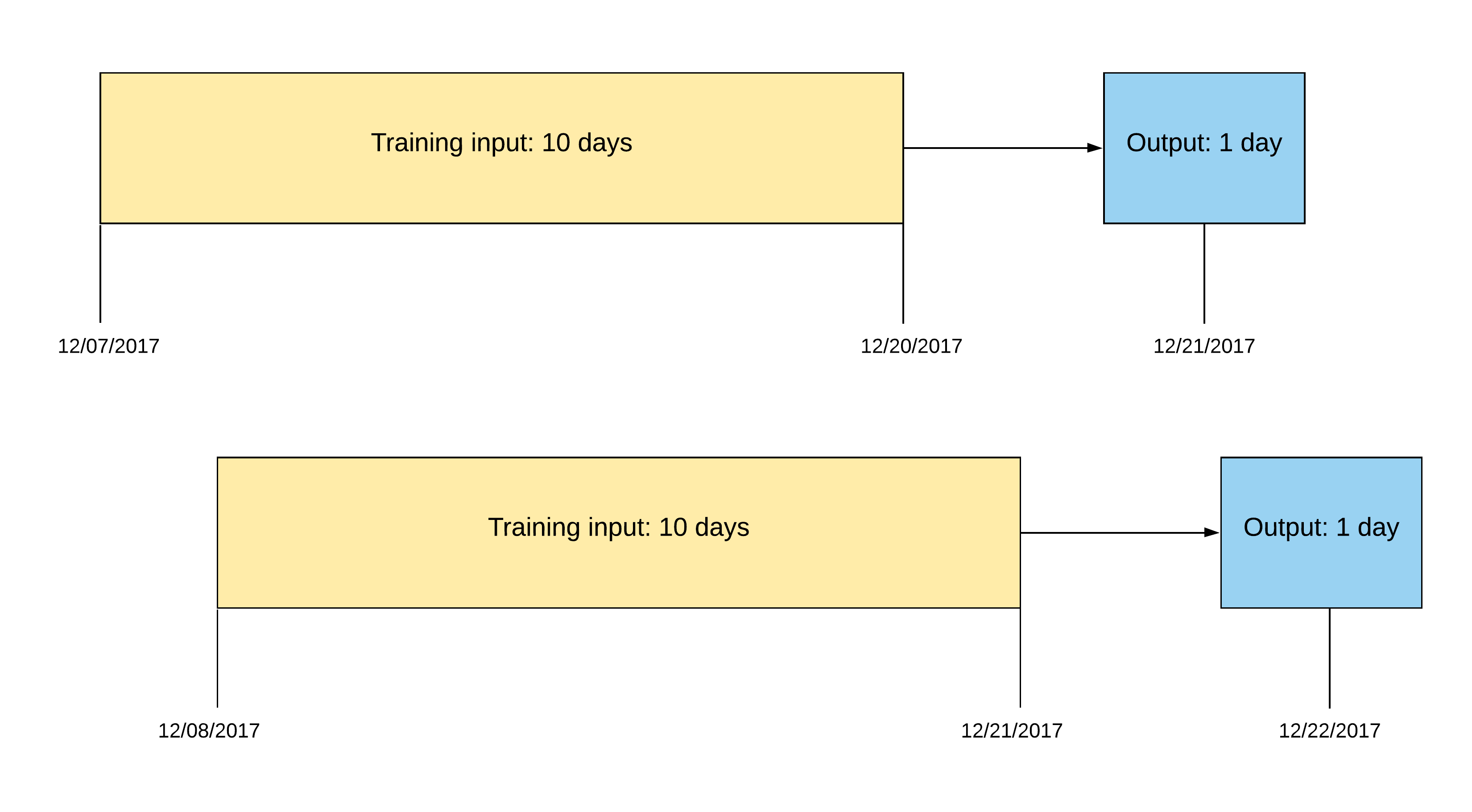}
\caption{\small Schematic diagram of the time step window.}
\label{fig:window_step}
\end{center}
\end{figure}

Since we are dealing with time series forecasting, a rolling window with fixed-size 10 is used to provide different time steps. As shown in Figure \ref{fig:window_step}, we are using the past 10 days' financial news from multiple sources and stock prices to predict the next day's stock price. The rolling window data are historical stock prices, and then we shift the rolling window by one day and add the next actual stock price to the last date of the rolling window to predict the next day's stock price and so forth. According to the length of the window, the training data is divided into 83-time steps. The validation data is divided into 9-time steps, and the test data divided is into 9-time steps.

% normalization
Normalization is a rescaling of the data from the original range; all scaled values are within the range between 0 and 1. Since the compound scores are numbers between -1 and 1, to avoid overfitting and improve accuracy, we rescaled the adjusted closing stock prices between 0 and 1. 

\section{Methodology}
\label{headings}

Stock price prediction is classified in the time-series category due to its unique characteristic, which means stock price prediction is a continual prediction following the direction of time. The most common techniques used for stock forecasting are statistics algorithms and economics models. However, the results coming out from there are not satisfactory because statistical algorithms and economics models cannot capture the stock movement's patterns. In Artificial Intelligence, the core techniques are pattern recognition using arithmetic calculations and sequences. RNNs utilize their internal memory to process variable-length sequences of input data, which makes them well-suited for time series forecasting\cite{Connor1994RecurrentNN}. In particular, LSTM and GRU are the first choices because they have been used future stock predictions successfully before. However, previous research has found that for a single LSTM network or GRU network, unless scrupulous parameter optimization is performed, data trained with a specific data set is likely to perform poorly on completely different time-series datasets. After extensive research and experiments, we have found that stack or combine multiple RNNs will provide a more accurate prediction compared to a single LSTM network\cite{krstanovic_paulheim_2017}. Therefore, we decide to deploy a blending ensemble learning model that combines LSTM and GRU to accomplish this difficult task.%Differences between LSTM and GRU
The main differences between LSTM and the GRU are the exposure of memory content inside the unit and how new information is processed by each unit. For the LSTM unit, the amount of memory content that is seen controlled by the output gate (not all of the content are exposed to other units; the output gate decides what information will be used in the next LSTM unit). The GRU unit exposes its full content to other units without any control. When LSTM receives new content, these new contents will be transported to the forget gate because the forget gate decides what information will be throw away or be kept before the computation process. However, the GRU does not have the forget gate; instead, the GRU utilizes the update gate to control the previous unit's information flow when computing the new candidate activation\cite{chung2014empirical}. Even though these two models are similar, but the way they process data and computation process steps are different. These differences might have an impact on weights when dealing with stock and news data. We have found that sometimes the LSTM prediction is more close to the actual stock price during our experiment, while other times the GRU prediction performs better. As we know some news contents may affect stock more and longer than usual, and other contents may affect stock in a short time. Due to its controlled exposure of the memory content in LSTM, it can filter out a lot of news contents. On the other hand, GRU can outperform LSTM both in terms of convergence in CPU time and in terms of parameter update and generalization\cite{chung2014empirical}. Thus, both models have their strengths and weaknesses, and in our design we hope to use both models with different parameters to complement each other in order to achieve the best prediction results.

\subsection{Ensemble Learning Techniques}

\begin{figure}[ht]
\begin{center}
\includegraphics[scale=0.30]{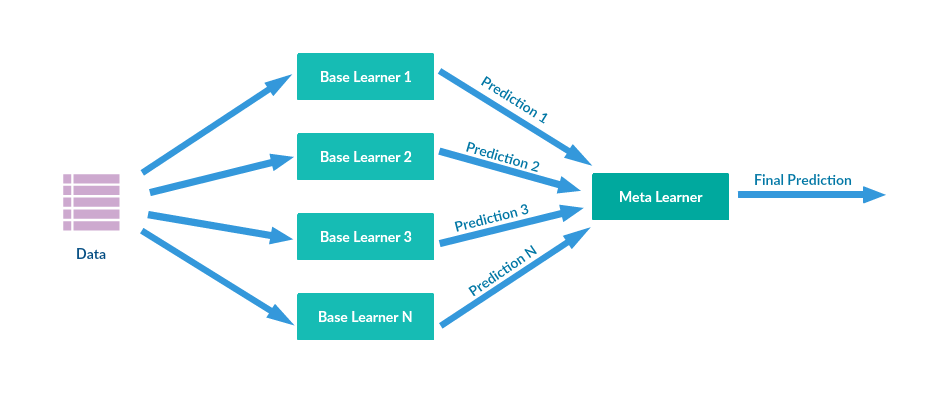}
\caption{\small source: https://www.kdnuggets.com/2019/01/ensemble-learning-5-main-approaches.html}
\label{fig:ensemble_graph}
\end{center}
\end{figure}

The ensemble learning method combines decisions from multiple sub-models to a new model and then to make the final output to improve the prediction accuracy or the overall performance  (See Figure \ref{fig:ensemble_graph}). There are many different ensemble learning models: Max Voting, Averaging, Weighted Average, Stacking, Blending, Bagging, Boosting, Adaptive Boosting (AdaBoost), Gradient Boosting Machine (GBM), eXtreme Gradient Boosting (XGB), etc\cite{10.1007/3-540-45014-9_1}. Different ensemble models hve different characteristics and can be used to solve different problems in a variety of domains. A simple example to describe the ensemble learning method is that compared with an individual's decision, a diverse group of people are more likely to make a better decision. The same principle applies to machine learning and deep learning models; a different set of models are more likely to perform a better comparison to a single model\cite{dietterich_2002} since each model has their own strength and they can complement each other to overcome their individual shortcomings. 

\subsection{Long Short-Term Memory Neural Networks}

Humans do not start their thinking from scratch every second. They understand each word based on their understanding of previous words. Hence, memory is important in recognition and traditional neural networks do not have this memory capability. The Long Short Term Memory (LSTM) is a special kind of recurrent neural network originally proposed by Hochreiter and Schmidhuber in 1997\cite{doi:10.1162/neco.1997.9.8.1735}. LSTM contains some memory cells, and can solve many time series tasks unsolvable by feed-forward networks or other machine learning models\cite{10.1007/978-1-4471-0219-9_20}. LSTM is very suitable and particularly successful in the time-series domain because LSTM can store important past information into the cell state and forget the unimportant information. LSTM has three gates that complete these complex tasks by adjusting input and stored information in the cell state. The first gate is the forget gate layer, which decides what information the unit will eliminate from the cell state. The forget gate equation is defined as

\[f_{t}=\sigma (W_{f}\cdot [h_{t-1},x_{t}]+b_{f}) \]

where $f_{t}$ represents the forget gate at the time step $t$; $\sigma$ represents a sigmoid function; $W_{f}$ represents the weights for the forget neurons; $h_{t-1}$ represents the output of the previous cell state at time step $t-1$; $x_{t}$ represents the input value at current time step; $b_{f}$ represents the biases for the forget gates.

There are two steps when LSTM decides what new information the unit will store in the cell state. The first step is the input gate layer, which determines which values will be updated. The second step is the tanh layer, which generates a new value-added to the present cell state. The equations are defined as

\[i_{t}=\sigma (W_{i}\cdot [h_{t-1},x_{t}]+b_{i})\]

\[\widetilde{C}_{t}=tanh(W_{C}\cdot [h_{t-1},x_{t}]+b_{C})\]

where $i_{t}$ represents the input gate at the time step $t$; $W_{i}$ represents the weights for the input neurons; $\widetilde{C}_{t}$ represents the candidate for the cell state at time step $t$; and $b_{i}$ and $b_{C}$ represents the biases for the respective gates. 

The last gate is the output gate layer, which determines what information will be output. The output gate equation is defined as

\[o_{t}=\sigma (W_{o}\cdot [h_{t-1},x_{t}]+b_{o})\]

where $o_{t}$ represents the output gate at the time step $t$; $W_{o}$ represents the weights for the output neurons; and $b_{o}$ represents the biases for the output gates\cite{doi:10.1162/neco.1997.9.8.1735}.

We implemented an LSTM which uses the past 10 days as the time window, and input data include adjusted closing stock price, four news sentiment compound scores to predict the next day's adjusted closing stock price. The details of the LSTM structures will be discussed in section 4.4.

\subsection{Gated Recurrent Unit Neural Networks}

A Gated Recurrent Unit (GRU) was deployed and proposed by Cho et al. in 2014\cite{cho2014properties} to solve the vanishing gradient problem of the traditional RNN by using an update gate and a reset gate. GRU is also a special kind of recurrent neural network that is very similar to LSTM; GRU has gating units that regulate the flow of information inside the unit. However, GRU does not have a separate memory cell, and that is one of the main differences between GRU and LSTM. The performance of the LSTM and GRU are equally matched in different test environments\cite{chung2014empirical}. However, GRU is computationally more efficient because GRU does not have to use a memory unit. Besides, GRU is more suitable and performs better when dealing with small datasets. The GRU's update gate decides how much of the past information needs to update before passing to the next step. The update gate equation is defined as

\[z_{t} = \sigma (W^{z}x_{t} + U^{z}h_{t-1})\]

where $z_{t}$ represents the update gate at time step $t$; $W^{z}$ represents the weights for the update gate; $x_{t}$ represents input at time step $t$; $h_{t-1}$ represents the holding values for the previous $t-1$ units; and $U^{z}$ represents the weights for the $h_{t-1}$.

The second principal component of the GRU is the reset gate. The reset gate decides how much of the past information needs to forget. The reset gate equation is defined as

\[r_{t} = \sigma (W^{r}x_{t} + U^{r}h_{t-1})\]

where $r_{t}$ represents the reset gate at time step $t$; $W^{r}$ represents the weights for the reset gate;
GRU can store and filter the information by utilizing the update gates and reset gates. The update and reset gates technique effectively eliminates the RNN vanishing gradient problem; they store relevant information in the memory cell and pass the values down to the next time steps of the network.

\subsection{Architecture Overview}

\begin{figure}[ht!]
\begin{center}
\includegraphics[scale=0.60]{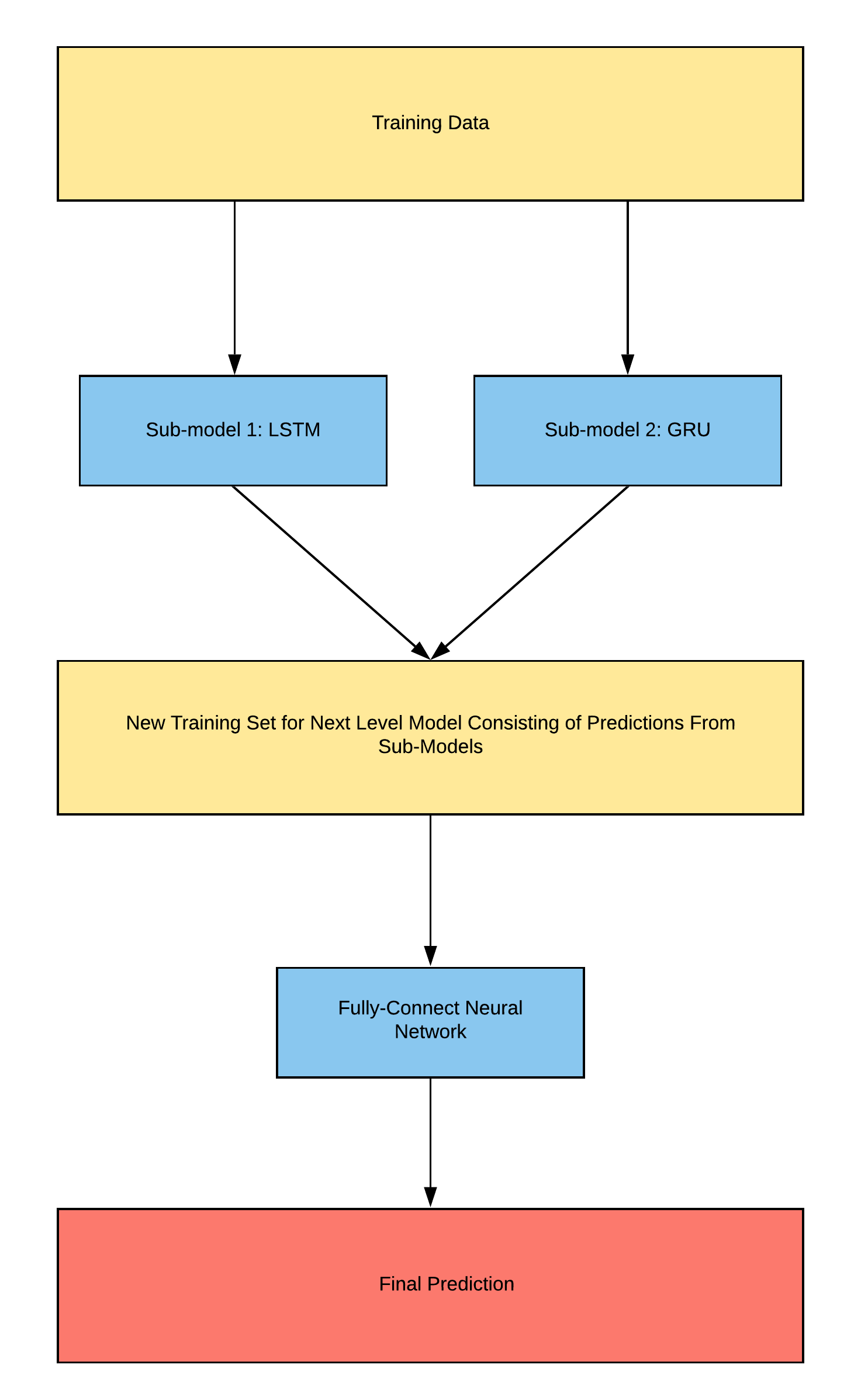}
\caption{\small Architecture Overview}
\label{fig:architecture}
\end{center}
\end{figure}

The concept of ensemble learning is to use different types of machine learning and deep learning models combined to make predictions or classifications. We deploy an ensemble model called the Blending Ensemble model; the overview of the architecture is shown in Figure \ref{fig:architecture}.

The Blending Ensemble model has two levels: the first level contains two RNNs; sub-model 1 is the LSTM, and sub-model 2 is the GRU model. 
%the purpose of having train, validation, test data
We already divid the dataset into three parts: training data (from 12/07/2017 to 04/09/2018), validation data (from 04/10/2018 to 05/04/2018), and test data (from 05/07/2018 to 06/01/2018). Each dataset is essential to train the Blending Ensemble model. The training data are used to train level 1 sub-models: the LSTM model and the GRU model. After the first phase training, we use the trained level 1 models to make predictions on the validation data, which is basically the level 2 model's training data. And the test data are used to make the final prediction and accuracy calculation.

First, we use the training data to train the sub-model 1: LSTM model. This LSTM model has only four layers, and each layer contains 50 neurons. We add 0.2 drops out for each hidden layer and train the model with 100 epochs. After we train the LSTM model, we input the validation dataset into the LSTM to make the first prediction with the validation dataset. The first prediction that the LSTM made using the validation called LSTM validation predictions. 

Secondly, we train the sub-model 2: the GRU model. The GRU model we build also contains four layers, and each layer has 50 neurons. We also add 0.2 drops out for each hidden layer and train the model with 100 epochs. The training process for the GRU model is the same as the LSTM. We input the training dataset into the GRU, and after we obtain a trained GRU model, we will input the validation dataset to make the GRU validation predictions. 

Once the LSTM validation prediction and GRU validation prediction are obtained, we combine them into a new training dataset in the form of $p \times m$ ( $p$ represents number of predictions and $m$ represents number of models). This new training data will be pass to the second level to train the meta-learner. The meta-learner is also called the second-level model. The meta-learner is a fully-connect neural network with three layers; the activation function for this model is the Rectified Linear Unit (ReLu). After the meta-learner is trained, the test dataset will be input into the sub-models again to produce intermediate test data for the meta-learner. Afterward, the meta-learner will use the intermediate test predictions from the sub-models to make the final predictions.

\section{Evaluation Metrics}

During the experiments, we mainly use four different evaluation metrics: mean square error (MSE), confusion matrix, mean prediction accuracy (MPA) and Movement Direction Accuracy (MDA) to evaluate the Blending Ensemble model. The MSE is a risk function that measures the average squared difference between the predicted values and the actual value. The MSE is defined as

\[ MSE = \frac{1}{n}\sum_{i=1}^{n} ( y_{i} - \widetilde{y}_{i})^{2} \]

where $n$ is the number of predictions, $y_{i}$ is the vector of observed values of the variable being predicted, with $\widetilde{y}_{i}$ being the predicted values. 
\begin{figure}[ht]
\begin{center}
\includegraphics[scale=0.8]{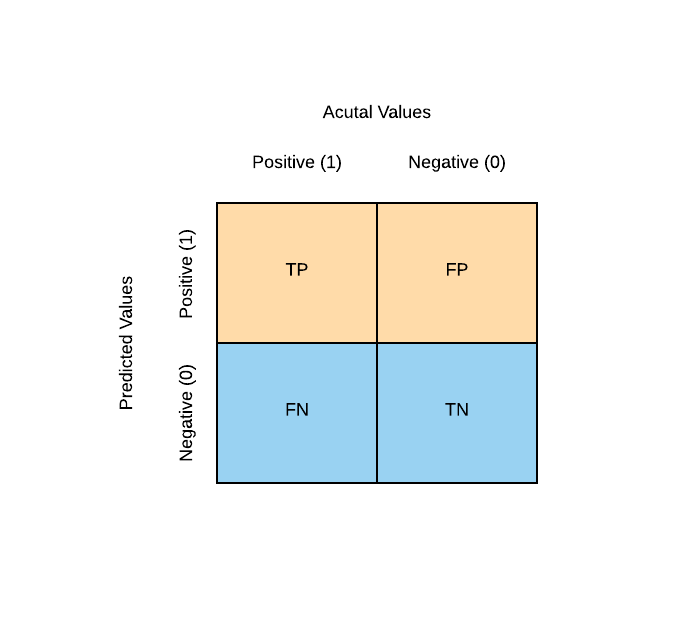}
\caption{\small Confusion Table}
\label{fig:confusion_table}
\end{center}
\end{figure}

The confusion matrix is usually used in statistical classification, also known as the error matrix. As shown in Figure  \ref{fig:confusion_table}, the confusion matrix is a unique table layout used to visualize the performance of an algorithm, classification scheme, or prediction model on a set of data where the actual values are known. The confusion matrix has different equations to measure the performance of the model. In this paper, we will use the Precision, Recall, and F1-Score to evaluate the the Blending Ensemble model. 

Precision indicates how precise the model is out of the positive predictions by calculating how many predicted positive is actually positive. The Precision is defined as

\[ Precision = \frac{TP}{ TP + FP} \]

where $TP$ represents the observation is positive, and the prediction is also positive. $FP$ represents the observation is positive, and the prediction is negative.

The Recall is defined as 
\[ Recall = \frac{TP}{ TP + FN} \]

where $FN$ represents the observation is positive, but the prediction is negative. Recall calculates the ratio of the actual positives the model captures through labeling it as positive. 
The Recall score indicates what percentage of the class is correctly recognized.

The F1-score is the combination of the Recall and Precision. F1-score uses the harmonic mean of Precision and Recall to compute the accuracy of the model where the score reaches its best value at 1 and worst at 0.

F1-score evaluation equation: 
$$F1-score = \left( {\frac{ Recall^{-1} + Precision^{-1} }{2}}\right)^{-1} = 2 * \frac{\left(Precision * Recall\right)}{\left(Precision + Recall\right)}\qquad$$

It penalizes the extreme values, which make it a better evaluation metric when dealing with imbalanced datasets. It also gives a better measure of the incorrectly classified cases than other metrics.

The next evaluation metric is the mean prediction accuracy (MPA) which is defined as 
\[ MPA_{t} = 1- \frac{l}{L}\sum_{{l}=1 }^{L}\frac{\left |  X_{t,{l} } - \hat{X}_{t,{l} } \right |}{X_{t,{l} }} \]

where $X$ represents the actual stock price, $\hat{X}$ represents the predicted stock price, ${l}$ represents the ${l}$-th stock, and $t$ represents the day\cite{li2019dplstm}.

Finally, the Movement Direction Accuracy (MDA) evaluates the percentage of correct prediction of stock movement directions (positive or negative) during the whole process. Movement Direction Accuracy (MDA) equation is define as:

\[Movement Direction Accuracy = \frac{Number\ of\ Correct\ Movement\ Predictions}{ Total\ Number\ of\ Movement\ Predictions} \]

\section{Experimental Results}

During the experiments, we use the predicted stock price to compare with the actual stock price and calculate the MSE and MPA values. We then use the predicted values to calculate the price fluctuation of the stock on the forecast day; if the predicted stock price increases, the output is 1, and if the predicted stock price decreases, the output is 0. These results will be used for the confusion matrix to calculate various measurements defined above. We first compare the Blending Ensemble model with the LSTM model, which is previous work widely used in the stock prediction domain, then we compare with another previous work called the DP-LSTM model. Based on the reported results in \cite{li2019dplstm}, the MSE is 198.75. However, we found an error in their code and after running the correct code, the actual MSE is 330.97. In the following comparisons, we will use the values from our experiments with the correct code. To further evaluate the Blending Ensemble model, we also deploy a GRU model that is very similar to the sub-model 2 that we were building as a test model. In addition, we also recorded the Averaging Ensemble model and Weighted Average Ensemble model prediction results to contrast with the Blending Ensemble model.

\begin{tabular}{|p{1.7cm}||p{1.5cm}|p{1.5cm}|p{1.5cm}|p{1.6cm}|p{1.6cm}|p{1.5cm}|}
 \hline
 \hline
 Evaluation Metrics&LSTM&DP-LSTM&GRU&Averaging Ensemble&Weighted Average Ensemble&\textbf{Blending Ensemble}\\
 \hline
 MSE&438.94&330.97&249.34&231.16&229.52&\textbf{186.32}\\
 \hline
 MPA&99.29\%& 99.48\%&99.57\%&99.57\%&99.57\%&\textbf{99.65\%}\\
 \hline
 Precision&25\%&20\%&40\%&25\%&40\%&\textbf{60\%}\\
 \hline
 Recall&25\% &25\%&50\%&25\%&50\%&\textbf{75\%}\\
 \hline
 F1-Score&25\%&22.22\% &44.44\%&25\%&44.44\%&\textbf{66.67\%}\\
 \hline
 MDA&33.33\%&22.22\% &44.44\%&33.33\%&33.33\%&\textbf{66.67\%}\\
 \hline
\end{tabular}

As shown in the table above, the Blending Ensemble model outperforms all other models in every category. The Blending Ensemble model has significant improvement in the MSE, Precision, Recall, and F1-Score evaluation categories. The MPA results are very similar; the Blending Ensemble model increased by 0.36\% compare with LSTM, increased 0.17\% compare with DP-LSTM and increased by 0.08\% compare with GRU, Averaging, Weighted Average Ensemble models. As the results have shown in the above table, the Blending Ensemble model improves the results in every evaluation metric compared with previous work and the test models.

% MSE comparison
\begin{figure}[ht]
\begin{center}
\includegraphics[scale=0.30]{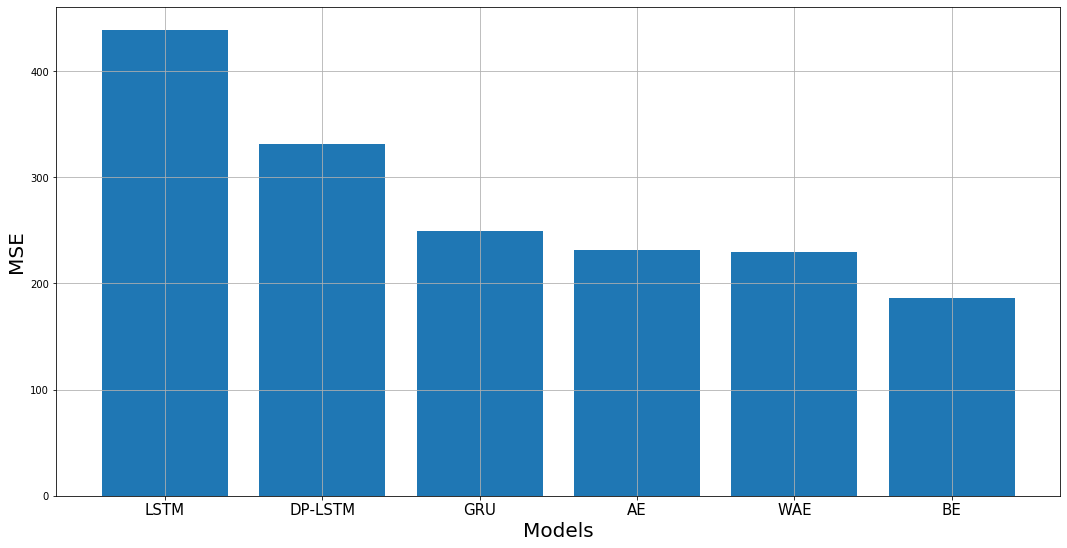}
\caption{\small Mean Square Error Comparison: AE represents Averaging Ensemble, WAE represents Weighted Average Ensemble, and BE stands for Blending Ensemble mode}
\label{fig:MSE_Comparsion}
\end{center}
\end{figure}

As shown in Figure \ref{fig:MSE_Comparsion}, the Blending Ensemble model reduces MSE from 438.94 to 186.32, up to 57.55\% reduction compared with baseline LSTM. When compared with the DP-LSTM model, the MSE reduces about 43.7\% and reduces MSE by 25.27\% compared with the test GRU model. The Averaging Ensemble and Weighted Average Ensemble MSE results are very close. The Blending Ensemble model reduced the MSE around about 20\%  compared with the Averaging Ensemble and Weighted Average Ensemble model.

% precision comparison 
\begin{figure}[h!]
\begin{center}
\includegraphics[scale=0.4]{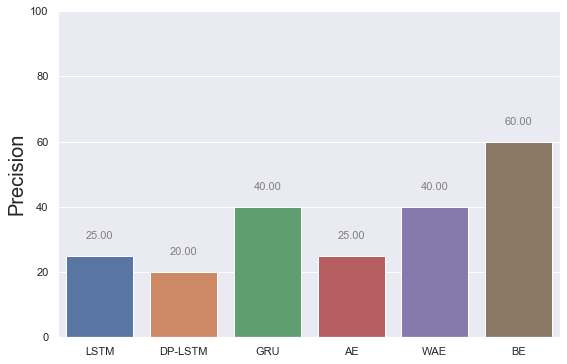}
\caption{\small Precision Comparison:
AE represents Averaging Ensemble, WAE represents Weighted Average Ensemble, and BE stands for Blending Ensemble model}
\label{fig:precision}
\end{center}
\end{figure}

In terms of Precision (See Figure \ref{fig:precision}), the Blending Ensemble model increased the accuracy percentage of at least 20\% compared with GRU and Weighted Average Ensemble model and up to 40\% when compared with the DP-LSTM model.

% recall comparison
\begin{figure}[h!]
\begin{center}
\includegraphics[scale=0.4]{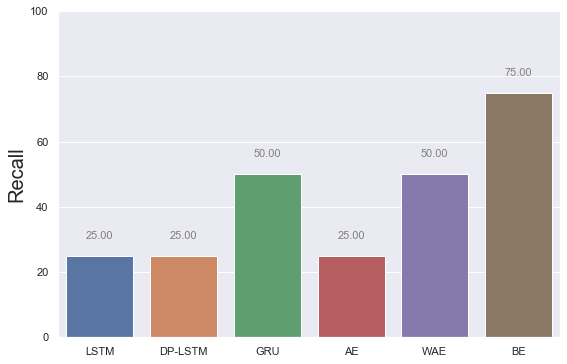}
\caption{\small Recall Comparison: AE represents Averaging Ensemble, WAE represents Weighted Average Ensemble, and BE stands for Blending Ensemble model}
\label{fig:recall}
\end{center}
\end{figure}

In terms of Recall (See Figure \ref{fig:recall}), the Blending Ensemble model increases the accuracy by 50\% compared with the LSTM, DP-LSTM and Averaging Ensemble model, and increases accuracy percentage by 25\% compared with the test GRU and Weighted Average Ensemble model.

% f1-score comparison
\begin{figure}[h!]
\begin{center}
\includegraphics[scale=0.4]{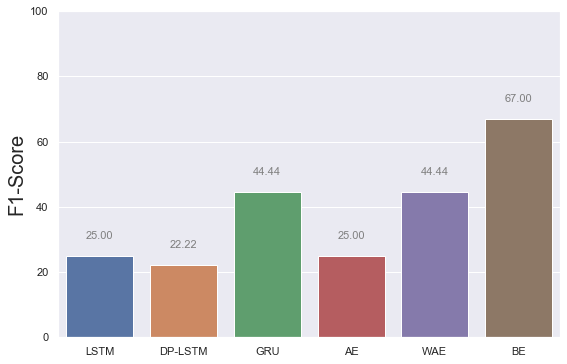}
\caption{\small F1-Score Comparison:
 AE represents Averaging Ensemble, WAE represents Weighted Average Ensemble, and BE stands for Blending Ensemble model}
\label{fig:f1-score}
\end{center}
\end{figure}

In F1-Score Comparison (See Figure \ref{fig:f1-score}), the Blending Ensemble model also significantly improved accuracy percentage. The Blending Ensemble model increased the accuracy percentage by 44.78\% compare with DP-LSTM, 42\% compare with LSTM and Averaging Ensemble model, and 22.56\% compare with GRU and Weighted Average Ensemble model. 

\begin{figure}[h!]
\begin{center}
\includegraphics[scale=0.4]{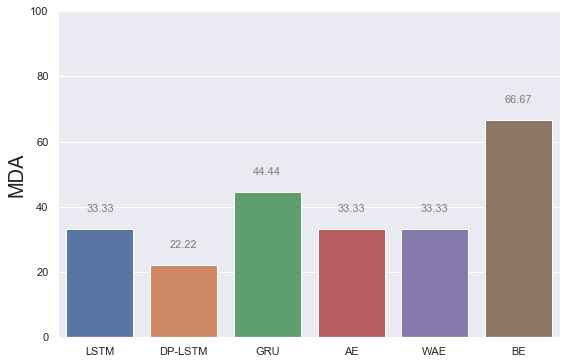}
\caption{\small MDA Comparison: AE represents Averaging Ensemble, WAE represents Weighted Average Ensemble, and BE stands for Blending Ensemble model}
\label{fig:MDA}
\end{center}
\end{figure}

Moreover, we also use the stock price fluctuation (positive or negative movement directions) to calculate the movement direction accuracy MDA of all the models in predicting future stock movement directions (See Figure \ref{fig:MDA}). 

% MDA comparison

On the test dataset predictions, the LSTM, Averaging Ensemble, and Weighted Average Ensemble model movement accuracy are around 33.33\%; DP-LSTM prediction average movement accuracy is around 22.22\%. The GRU model prediction average movement accuracy is around 44.44\%, and the Blending Ensemble model achieves approximately 66.67\% movement accuracy. 

% entire data set graph included prediction results
\begin{figure}[h!]
\begin{center}
\includegraphics[scale=0.24]{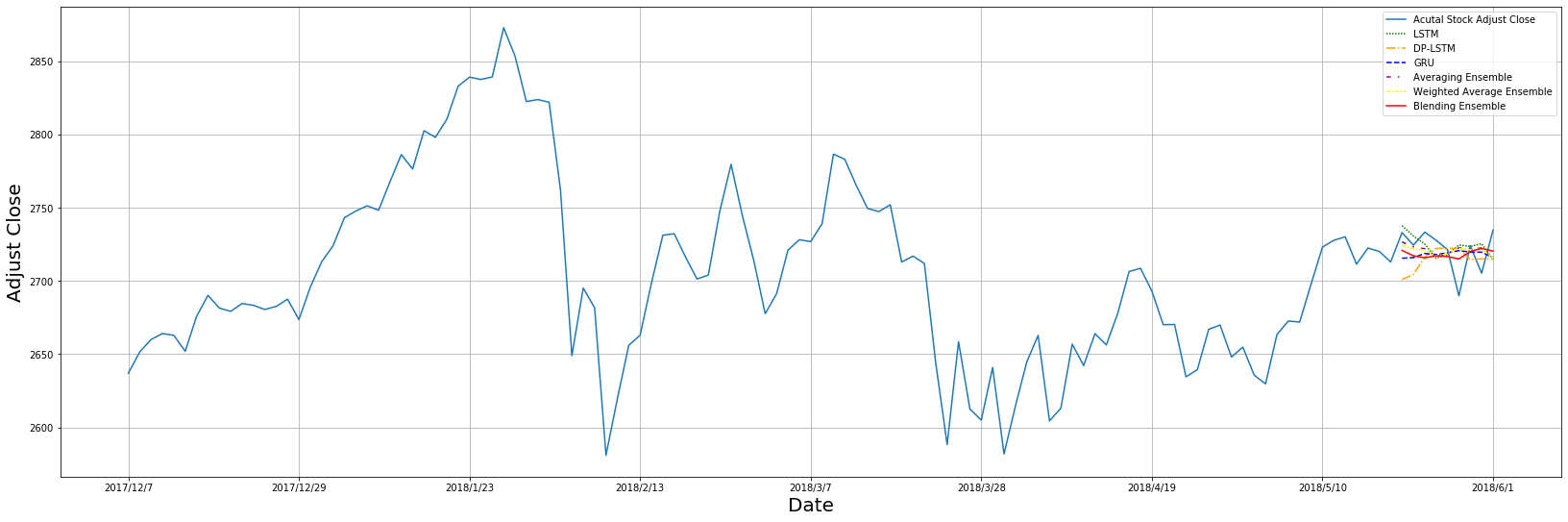}
\caption{\small Prediction result of the Blending Ensemble Learning model compare with LSTM, DP-LSTM, GRU test model, Averaging Ensemble, and Weighted Average Ensemble model}
\label{fig:whole_graph}
\end{center}
\end{figure}

In Figure \ref{fig:whole_graph}, the plot shown the entire data included training data, validation data, and testing data. As we can see, the stock price is very unsteady, and the stock's float is very large. However, we can see the prediction results of the Blending Ensemble model are more closer to the actual stock, and the pattern of the prediction line is more identical to the actual stock. 

\section{Conclusion}

This paper proposes a novel deep learning architecture to predict stock prices by combines multiple recurrent neural networks to form a blending ensemble learning model. The Blending Ensemble model captures the changes in time series influence on the stock price. To evaluate the effectiveness of the method, we have conducted experiments on different recurrent neural network models and set up the same environment and configures to compare with the Blending Ensemble model. Several evaluation metrics like MSE, MPA, Precision, Recall, F1-score and MDA are used to evaluate the performance of the Blending Ensemble model. Our experimental results show clearly that the Blending Ensemble model outperforms all previous methods in every category. According to one of the global tech news organization NikolaNEWS, if a machine learning or deep learning model can reach 60\% accuracy on predicting stock movement directions, it can deliver solid returns\cite{adusumilli}. Our model can reach up to 67\% in MDA on the hold-out test dataset, which strongly suggests that using deep learning models to analyze and predict future stock price or movement direction can be a valuable source to stock investment companies and individual investors. For other performance metrics, the Blending Ensemble model achieves up to 57.55\% improvement in MSE; In predicting the future direction of the stock market, the Blending Ensemble model improves Precision accuracy by 40\%, Recall accuracy percentage by 50\%, and F1-Score by 44.78\%, and Movement Direction Accuracy by 33.34\%, compared with the best results in the literature. These results demonstrate that our novel blend multiple different recurrent neural networks can significantly improve the previous best model's robustness and prediction results. Moreover, our findings open doors for more sophisticated ensemble deep learning models and using more complex data sources for stock prediction in the future. It is our hope that these new models will truly better assist stock investors in making the correct decision in a real world situation. 

\section{Future Research}
We believe that there is a lot of improvement space over the current blending ensemble model and input data sources. We may use many different ways to improve our current work. Below are possible future research directions.

1. There is a good chance that the current results could be improved by fine-tuning the hyperparameters, increasing the size of the training dataset, and considering other data sources such as the 10-K annual report. 

2. Understanding the mechanisms of our prediction can provide more insights into our prediction results. We plan to understand our prediction better through rule generation\cite{He2006RuleGF} and use other machine learning technologies such as Clustering SVM\cite{Zhong2007ClusteringSV} and Clustering Deep Learning \cite{zhong2020} to improve our results. 

3. We also plan to expand the current model by adding LSTM with attention and possibly combining more models to mining different data sources to achieve better prediction results. Especially in level 1, we may employ more parallel models to complement each other.

4. We would like to deploy a reinforcement learning model as the second level model to explore the area of stock market prediction further. Reinforcement learning is believed to get an optimal policy for a specific problem so that the reward or profit obtained under this strategy could be a better choice. The policy is actually a series of actions that are basically sequential data\cite{Sutton1998ReinforcementLA}. 

5. Also, fuzzy logic systems have been used in many applications, such as wireless network routing\cite{Liu2005AnAG}. We will introduce fuzzy deep learning into our learning and prediction models in the future\cite{8928970} since many news items are fuzzy in terms of their positive or negative impacts. 

6. We may also dynamically change the historical window size based on the type of news. Some news has long lasting impact such as housing costs, while others live a very short life such as a sudden disaster. We even can have different window sizes for different news types. Studying which news type has a long term effect and how long it lasts is probably more a psychological study than a computer science problem. But combining these discoveries in psychology, economy and political science may help our prediction do a better job. 

As Artificial Intelligence becomes more powerful, computer scientists are also constantly developing new models to analyze and predict the stock market, hoping to provide more reliable and more precise stock information to investors. Although our study is preliminary, it is a good start for more interdisciplinary research in this exciting area.

\section{Acknowledgement}
%The author thank Professor Yi Pan for his valuable discussions, comments and support for this research. 

The authors would like to thank Yinchuan Li for providing the datasets for this research and for his generous support and Sean Cao for his helpful comments.

\bibliography{reference}
\bibliographystyle{IEEEtran}

\end{document}